\documentclass{article}
\usepackage{amsfonts}
\usepackage{amsmath}

\begin{document}

\title{Variable selection based on entropic criterion and its application to
the debris-flow triggering}
\author{Chien-chih Chen$^{1}$, Chih-Yuan Tseng$^{2}$ and Jia-Jyun Dong$^{3}$ 
\\
%EndAName
$^1$ Institute of Geophysics, $^2$ Department of Physics, and \\
$^3$ Institute of Applied Geology\\
National Central University, Jhongli, Taiwan 320 }
\date{}
\maketitle

\begin{abstract}
We propose a new data analyzing scheme, the method of minimum entropy
analysis (MEA), in this paper. New MEA provides a quantitative criterion to
select relevant variables for modeling the physical system interested. Such
method can be easily extended to various geophysical/geological data
analysis, where many relevant or irrelevant available measurements may
obscure the understanding of the highly complicated physical system like the
triggering of debris-flows. After demonstrating and testing the MEA method,
we apply this method to a dataset of debris-flow occurrences in Taiwan and
successfully find out three relevant variables, i.e. the hydrological form
factor, numbers and areas of landslides, to the triggering of observed
debris-flow events due to the 1996 Typhoon Herb.
\end{abstract}

\section{Introduction}

Most geophysical/geological problems, e.g. the trigging of debris-flows or
earthquakes, are so complicated that many observed and/or unobserved
variables have their obscure contributions to the geophysical/geological
systems \cite{Rundle00}. From the viewpoint of practical experiment setting,
scientists quite often encounter a problem of variable selection to choose
relevant measurement regarding their own physical systems. For instance,
three categories of variables describing three aspects of topography,
geology and hydrology, are usually used in the geographic information system
(GIS) to assess the hazard potential of debris-flows (e.g. \cite{LinPS02}; 
\cite{Rupter}). Although some consensus could be reached for the problem of
debris-flow trigging, the variables measured could be much different among
different research groups \cite{Wieczorek}. Therefore, when lots of
measurement could be probably made and available to use, we are forced to
face a fundamental question of which variables are relevant to describe a
highly complex physical system like the debris-flow system.

To answer the abovementioned question, we in this study introduce a new data
analyzing scheme, i.e. the minimum entropy criterion (\cite{Tseng06}; \cite%
{Tseng062}), to the problem of selecting the variables which dominate the
debris-flow occurrence. We first present the principle of minimum entropy
analysis (MEA) and verify its result when applying to a geological example
extracted from the textbook of Davis \cite{Davis02}. Then, in Sec. 3, we
demonstrate the application of MEA to an observed debris-flow dataset \cite%
{LinCW00}, consisting of the binary outcome (the response) of debris-flow
occurrences and measurement (the covariates) of some topographic, geologic
and hydrologic variables. Conclusion will be given at the end of this paper.

\section{Minimum entropy rule for variable selection}

\subsection{Principle of minimum entropy analysis}

Model selection in data processing is usually achieved by ranking models
according to the increasing order of \textit{preference}. Several methods
such as P-values, Bayesian approaches and Kullback-Leibler distance method,
etc., are some popular examples to provide pertinent selection criteria (%
\cite{Tseng06}; \cite{Weiss}; \cite{Raftery97}; \cite{Forbes03}; \cite{Dupuis03}). 
Tseng \cite{Tseng06} reviews those methods and suggests an
entropy-based criterion as the selecting preference of models.

Principle of maximum entropy proposed by Jaynes (\cite{Jaynes57a}, \cite%
{Jaynes57b}, \cite{Jaynes03}) is recognized as a tool to assign a
probability distribution to a system. This tool involves the use of a unique
functional form of entropy $S\left[ P\right] =-\sum_{x}P\left( x\right) \ln
P\left( x\right) $, where x denotes states of the model and P is the
probability density function. It is uniquely determined through Shannon's
axiomatic approach. Since Jaynes's work, this tool was further extended to
become an inductive inference tool for information processing, for updating
the probability distribution of a system according to information in hand (%
\cite{Skilling}; \cite{Caticha04}). Similarly, when a reference density
function $m(x)$ is available, one can also show that the \textit{relative}
entropy involves another unique functional form, $S\left[ P\left\vert
m\right. \right] =-\sum_{x}P\left( x\right) \ln P\left( x\right) /m\left(
x\right) $ \cite{Tseng06}.

Tseng \cite{Tseng06} has shown that the relative entropy uniquely determines
the preference for model selection. Suppose a family of models is given by
probability distributions $\left\{ P^{m}\left( x\right) \right\} $, where m
labels the model. The preference given by the relative entropy of model $%
P^{m}\left( x\right) $ and a reference measure $\mu \left( x\right) $,

\begin{equation}
S\left[ P^{m}\left\vert \mu \right. \right] =-\sum_{x}P^{m}\left( x\right)
\ln P^{m}\left( x\right) /\mu \left( x\right) \mbox{ ,}  
\label{S[P|mu]}
\end{equation}%
is a scalar value. Such scalar relative entropy measures differences between
model $P^{m}\left( x\right) $\ and a reference measure $\mu \left( x\right) $
\cite{Tseng06}. Maximizing the relative entropy $S\left[ P^{m}\left\vert \mu
\right. \right] $ indicates $P^{m}\left( x\right) $ to equal to the
reference measure $\mu \left( x\right) $.

If the reference measure is chosen to be the distribution 
$P_{\text{real}}\left( x\right)$ that is believed to be able to interpret the system
correctly, maximizing $S\left[ P^{m}\left\vert P_{\text{real}}\right. \right]
$ indicates that the model $P^{m}\left( x\right) $ is the most preferable.
Unfortunately, the real distribution is usually difficult to be practically
determined. Tseng \cite{Tseng06} proposes to rank models according to the
relative entropy $S\left[ P^{m}\left\vert \mu \right. \right] $ with the
reference measure $\mu \left( x\right) $ being set to a uniform probability
distribution $P_{\text{uni}}\left( x\right) $. Since a uniform distribution
does not carry any information about the system, maximum $S\left[
P^{m}\left\vert P_{\text{uni}}\right. \right] $\ indicates that $P^{m}\left(
x\right) $ is identical to $P_{\text{uni}}$ and the model $P^{m}\left(
x\right) $ carries no information about the system at all. On the other
hand, when a model $P^{m}\left( x\right) $ is codified with more
information, $P^{m}\left( x\right) $ differs from the uniform distribution
more. Thus, decreasing the relative entropy $S\left[ P^{m}\left\vert P_{%
\text{uni}}\right. \right] $ should provides same preference of different
models given by increasing $S\left[ P^{m}\left\vert P_{\text{real}}\right. %
\right] $.

In the case of variable selection, let's suppose a regression model $P\left( 
\overrightarrow{x}\right) $ associated with N variables $\overrightarrow{x}%
=\left\{ x_{1},x_{2},\cdots x_{N}\right\} $ is given to reveal the behavior
of an unknown system from experimental measurements. For example, the logit
model is often used for a system with the binary outcome (\cite{Johnson99}; 
\cite{Dupuis03}; \cite{Rupter}). Note that those variables $\overrightarrow{x%
}$ are usually assessed according to experiments (observations) and may or
may not denote crucial characteristics of the system interested. Besides,
they may be correlated to each other. Our question, then, is that, after
modeling an unknown system with different combinations of variables, which
ones play the most important roles allowing the model to pertinently
interpret the system. Namely, what is the preference of those variables?
This is basically the same question addressed in model selection by Tseng 
\cite{Tseng06}.

Suppose that a full model defined by $P_{\text{full}}\left( \overrightarrow{x%
}\right) $ is the model containing all N variables available from
experiments. Since given a set of N variables, there will be 2N - 2
combinations (subsets) of variables $\overrightarrow{x}_{s_{i}}\in 
\overrightarrow{x}$ . Each subset forms a submodel $P_{s_{i}}\left( 
\overrightarrow{x}_{s_{i}}\right) $. According to Eq. (\ref{S[P|mu]}) with $%
P^{m}\left( x\right) $ being replaced by $P_{s_{i}}\left( \overrightarrow{x}%
_{s_{i}}\right) $ and $\mu \left( x\right) $ being given by a uniform
distribution, increasing ranking order of the preference for these submodels
is given by decreasing the relative entropy

\begin{equation}
S\left[ P_{s_{i}}\left\vert P_{\text{uni}}\right. \right] =-\sum_{%
\overrightarrow{x}_{s_{i}}\in \overrightarrow{x}}P_{s_{i}}\left( 
\overrightarrow{x}_{s_{i}}\right) \ln \frac{P_{s_{i}}\left( \overrightarrow{x%
}_{s_{i}}\right) }{P_{\text{uni}}}=S\left[ P_{s_{i}}\right] +\ln P_{\text{uni%
}}\text{ ,}  \label{S[Ps|Puni]}
\end{equation}%
where the submode $P_{s_{i}}\left( \overrightarrow{x}_{s_{i}}\right) $l
contains $n_{i}$ variables and $S\left[ P_{s_{i}}\right] =-\sum_{%
\overrightarrow{x}_{s_{i}}\in \overrightarrow{x}}P_{s_{i}}\left( 
\overrightarrow{x}_{s_{i}}\right) \ln P_{s_{i}}\left( \overrightarrow{x}%
_{s_{i}}\right) $. Since $\ln P_{\text{uni}}$ is a constant, ranking order
given by decreasing $S\left[ P_{s_{i}}\left\vert P_{\text{uni}}\right. %
\right] $ is identical to that one from decreasing $S\left[ P_{s_{i}}\right] 
$. After determining the ranking preference of submodels, the selection of
variables then can be made from analysis of those submodels thus ranked. The
detailed process will be illustrated in the following.

\subsection{Demonstration and verification of minimum entropy analysis}

We test our data processing procedure of the MEA method with an example of
sample classification extracted from the book of \cite{Davis02}. Table 1
contains the results of brine analyses for oil-field waters from three
groups of carbonate units in Texas and Oklahoma. Brines recovered during
drillstem tests of wells may have relict compositional characteristics that
provide clues to the origin or depositional environment of their source
rocks. The first column in Table 1 denotes the brine samples belonging or
not belonging to some specific carbonate unit (Grayburg Dolomite, briefly in
\textquotedblleft Unit G\textquotedblright\ here), while the rest are the
percentages of six chemical ions. Davis \cite{Davis02} applies the
discriminant function analysis (DFA) to these six multivariate measurements
for finding a projection, i.e. a linear combinations of measurements,
allowing various categories of samples to be distinguished. The first
discriminant function thus calculated is (-0.3765, -0.0468, 0.0112, -0.0148,
-0.0174, -0.0110)$\cdot$(HCO$_{\text{3}}$, SO$_{\text{4}}$, Cl, Ca, Mg, Na)$%
^{\text{T}}$, which can clearly separates samples from Unit G and other
units. Note that the weighting factors in the first discriminant function
for variables of HCO$_{\text{3}}$ and SO$_{\text{4}}$, i.e. -0.3765 and
-0.0468, represent the first two largest factors in magnitude among six,
thus indicating these two variables play the most dominant effect in
classification.

%%%%%%%%%%%%%%%%%%%%%%%%%%%%%%%%%%%%%%%%%%%%%%%%%%%%%%%%%%%%%%%%
\vspace{-8pt} 
\begin{table}[h]
\caption{\baselineskip=8pt{\protect\footnotesize \textsf{Chemical analyses of brines 
(in ppm) recovered from drillstem tests of three carbonate rock 
units (Ellenburger Dolomite, Grayburg Dolomite = Unit G, Viola Limestone) in 
Texas and Oklahoma. Adapted from Davis \cite{Davis02}. }}}
\label{carbonate}%%%%%%%% {Carbonate} %%%%%%%%%%%%%%%%%
\centering
\begin{tabular}{|c|c|c|c|c|c|c|}
\hline
Unit G & HCO$_{3}$ & SO$_{4}$ & Cl & Ca & Mg & Na \\ \hline
N & 10.4 & 30 & 967.1 & 95.9 & 53.7 & 857.7 \\ \hline
N & 6.2 & 29.6 & 1174.9 & 111.7 & 43.9 & 1054.7 \\ \hline
N & 2.1 & 11.4 & 2387.1 & 348.3 & 119.3 & 1932.4 \\ \hline
N & 8.5 & 22.5 & 2186.1 & 339.6 & 73.6 & 1803.4 \\ \hline
N & 6.7 & 32.8 & 2015.5 & 287.6 & 75.1 & 1691.8 \\ \hline
N & 3.8 & 18.9 & 2175.8 & 340.4 & 63.8 & 1793.9 \\ \hline
N & 1.5 & 16.5 & 2367 & 412 & 95.8 & 1872.5 \\ \hline
Y & 25.6 & 0 & 134.7 & 12.7 & 7.1 & 134.7 \\ \hline
Y & 12 & 104.6 & 3163.8 & 95.6 & 90.1 & 3093.9 \\ \hline
Y & 9 & 104 & 1342.6 & 104.9 & 160.2 & 1190.1 \\ \hline
Y & 13.7 & 103.3 & 2151.6 & 103.7 & 70 & 2054.6 \\ \hline
Y & 16.6 & 92.3 & 905.1 & 91.5 & 50.9 & 871.4 \\ \hline
Y & 14.1 & 80.1 & 554.8 & 118.9 & 62.3 & 472.4 \\ \hline
N & 1.3 & 10.4 & 3399.5 & 532.3 & 235.6 & 2642.5 \\ \hline
N & 3.6 & 5.2 & 974.5 & 147.5 & 69 & 768.1 \\ \hline
N & 0.8 & 9.8 & 1430.2 & 295.7 & 118.4 & 1027.1 \\ \hline
N & 1.8 & 25.6 & 183.2 & 35.4 & 13.5 & 161.5 \\ \hline
N & 8.8 & 3.4 & 289.9 & 32.8 & 22.4 & 225.2 \\ \hline
N & 6.3 & 16.7 & 360.9 & 41.9 & 24 & 318.1 \\ \hline
\end{tabular}%
\end{table}
%%%%%%%%%%%%%%%%%%%%%%%%%%%%%%

Can we identify relevant variables in the problem of determining the
category of samples in Table 1, by means of our entropy-based procedure?

Let's consider the response to be the binary outcome belonging
(\textquotedblleft Y\textquotedblright\ or \textquotedblleft
1\textquotedblright ) or not belonging (\textquotedblleft
N\textquotedblright\ or \textquotedblleft 0\textquotedblright ) to Unit G
and the covariates those percentages of six chemical ions in Table 1. We can
apply the logit model (\cite{Dupuis03}; \cite{Johnson99})%
\begin{equation}
R\left( \overrightarrow{x}\right) =\frac{\exp \sum_{i=1}^{N}\beta _{i}x_{i}}{%
\exp \sum_{i=1}^{N}\beta _{i}x_{i}+1}  \label{Logit model}
\end{equation}%
to relate the response to the covariates. Normalizing Eq. (\ref{Logit model}%
), probability distribution of the response for a given subset of all 6
variables is%
\begin{equation}
P\left( \overrightarrow{x}\right) =R\left( \overrightarrow{x}\right) /Z=%
\frac{1}{Z}\frac{\exp \sum_{i=1}^{N}\beta _{i}x_{i}}{\exp
\sum_{i=1}^{N}\beta _{i}x_{i}+1}  \label{P=R/z}
\end{equation}%
where $Z=\sum_{\overrightarrow{x}}\frac{\exp \sum_{i=1}^{N}\beta _{i}x_{i}}{%
\exp \sum_{i=1}^{N}\beta _{i}x_{i}+1}$ is the normalization constant. Note
that coefficients $\beta _{i}$ could be determined through fitting the logit
model to experimental measurements by the maximum likelihood estimation \cite%
{Johnson99}. Thus, the entropy of $P\left( \overrightarrow{x}_{s_{i}}\right) 
$, i.e. Eq. (\ref{S[Ps|Puni]}), with different subsets of variables $%
\overrightarrow{x}_{s_{i}}\in \overrightarrow{x}$ gives the ranking order of
different submodels $P\left( \overrightarrow{x}_{s_{i}}\right) $ defined by
Eq. (\ref{P=R/z}) with $\overrightarrow{x}$ being replaced by $%
\overrightarrow{x}_{s_{i}}$.

In the example of brine data there are 62 submodels. We found 16 submodels
among those 62 to have the minimum entropy value of \symbol{126}1.7918 as
shown in Table 2, while the rest of the submodels have the entropy larger
than 2. The MEA suggests that these 16 submodels to be the most preferable.
Yet due to the intrinsically finite precision of measured data, we can not
distinguish the preferences of these 16 submodels further. Tackling the
issue of entropy resolution resulted from the intrinsic measurement
precision there are many possible ways (e.g. \cite{Dupuis03}) to determine
the most dominate variables in this example. Here we simply count the
frequencies of six variables appeared in these 16 submodels. It turns out
that the frequencies for variables of HCO$_{\text{3}}$ and SO$_{\text{4}}$
are 16 and 15, respectively, and 8 for the rest of variables. This result
suggests that the ability of interpreting the experimental measurements by
the logit model is strongly dominated by simultaneously associating
variables of HCO$_{\text{3}}$ and SO$_{\text{4}}$ in the data. And, such
result is much consistent with the DFA. Comparing both results from the DFA
and the MEA procedures improves the understanding and enhances the
confidence in our entropy-based technique.

%%%%%%%%%%%%%%%%%%%%%%%%%%%%%%%%%%%%%%%%%%%%%%%%%%%%%%%%%%%%%%%%
\vspace{-8pt} 
\begin{table}[h]
\caption{\baselineskip=8pt{\protect\footnotesize \textsf{Entropy (S) for 
sixteen submodels with different combinations of six variables in 
Table 1 (A = HCO3, B = SO4, C = Cl, D = Ca, E = Mg, and F = Na). "1" or "0" 
denotes the variable selected or not selected in each submodel.}}}
\label{table2}
\centering%
\begin{tabular}{|cccccc|c|}
\hline
A & B & C & D & E &F & S \\ \hline
1&	1&	1&	0&	1&	1&	1.79183823 \\ \hline
1&	1&	1&	1&	0&	1&	1.79183829\\ \hline
1&	1&	0&	1&	1&	1&	1.79183836\\ \hline
1&	1&	1&	1&	1&	0&	1.79183836\\ \hline
1&	1&	0&	1&	1&	0&	1.79184075\\ \hline
1&	1&	0&	0&	1&	1&	1.79184177\\ \hline
1&	1&	1&	0&	1&	0&	1.79184215\\ \hline
1&	1&	1&	0&	0&	1&	1.79184241\\ \hline
1&	1&	0&	0&	1&	0&	1.79184396\\ \hline
1&	1&	0&	1&	0&	1&	1.79184653\\ \hline
1&	1&	1&	1&	0&	0&	1.79184701\\ \hline
1&	0&	1&	1&	1&	1&	1.79184888\\ \hline
1&	1&	0&	1&      0&      0&	1.79184968\\ \hline
1&	1&	1&	0&	0&	0&	1.79185471\\ \hline
1&	1&	0&	0&	0&	1&	1.79185668\\ \hline
1&	1&	0&	0&	0&	0&	1.79185738\\ \hline

\end{tabular}%
\end{table}
%%%%%%%%%%%%%%%%%%%%%%%%%%%%%%

\section{Application of minimum entropy analysis to the debris-flow trigging}

Taiwan located at an active convergent plate boundary is an island with
rugged topography and severe erosion. During heavy rainfalls brought by
typhoons, the occurrence of debris-flows often results in enormous damage of
life and buildings (\cite{LinPS02}, \cite{LinCW00}, \cite%
{LinCW03}, \cite{ChenJD97}, and \cite{Jan05}). There are absolutely many factors intricately
affecting the occurrence of the debris-flows and various field measurement
has been conducted to assess the occurrence potential of debris-flows in
Taiwan (\cite{LinPS02}, \cite{Wieczorek}, \cite{LinCW00}, \cite{LinCW03}, 
\cite{Jan05}, \cite{Chen01}, and \cite{ChenJD00}). So, can we figure out the
observations relevant to the trigging of debris-flows by means of our MEA
procedure? To preliminarily apply the MEA method, we have used a relatively
small dataset documenting the occurrence of debris-flows (Table 3) in the
Hsinyi area of Nantou County, Central Taiwan, during the 1996 Typhoon Herb 
\cite{LinCW00}. The related variables including gully lengths (Le), areas of
drainage basin with slope $> 15^{o}$ (Ad), form factor (Ff =
Ad/Le$^{2}$), and numbers (Nl) and areas (Al) of landslides, which implicitly
reflect the topographic, geologic and hydrologic characteristics of examined
gullies, are listed in Table 3. For the detailed description of field
observations, please refer to the paper of Lin et al. \cite{LinCW00}.

%%%%%%%%%%%%%%%%%%%%%%%%%%%%%%%%%%%%%%%%%%%%%%%%%%%%%%%%%%%%%%%%
\vspace{-8pt} 
\begin{table}[h]
\caption{\baselineskip=8pt{\protect\footnotesize \textsf{Debris-flow occurrences 
of 22 creeks (the 1st column) during the 1996 Typhoon Herb in Hsin-Yi area of 
the Nantou County, Central Taiwan, together with their corresponding characteristics 
including gully lengths (Le), areas of drainage basin with slope $> 15^{o}$ (Ad), form 
factor (Ff = Ad/Le$^{2}$) and numbers (Nl) and areas (Al) of landslides. 
Adapted from Lin et al. \cite{LinPS02}.}}}
\label{table3} 
\centering%
\begin{tabular}{|c|c|c|c|c|c|}
\hline
Occurrence&	Le [m]&	Ad [km$^{2}$]&	Ff&	Nl&	Al [km$^{2}$]\\ \hline
No&	1505&	0.86&	0.3797&	0&	0\\ \hline
Yes&	1876&	1.27&	0.3609&	3&	10.9\\ \hline
Yes&	1640&	0.35&	0.1301&	2&	4.5\\ \hline
Yes&	1560&	0.57&	0.2342&	3&	3.4\\ \hline
Yes&	2158&	1.82&	0.3908&	5&	3.7\\ \hline
Yes&	1035&	1.82&	1.6990&	1&	7.8\\ \hline
Yes&	582&	3.4&	10.0377&	5&	3.7\\ \hline
Yes&	2445&	3.4&	0.5687&	4&	4.4\\ \hline
Yes&	2685&	3.5&	0.4855&	9&	8.4\\ \hline
No&	2350&	1.97&	0.3567&	0&	0\\ \hline
No&	142&	1.15&	57.0323&	4&	2.7\\ \hline
No&	1349&	1.55&	0.8517&	4&	2.9\\ \hline
No&	1337&	0.74&	0.4140&	4&	1.2\\ \hline
No&	911&	0.72&	0.8676&	3&	0.8\\ \hline
Yes&	2048&	0.78&	0.1860&	5&	0.086\\ \hline
Yes&	2960&	2.18&	0.2488&	6&	0.226\\ \hline
No&	2010&	1.65&	0.4084&	6&	0.061\\ \hline
Yes&	675&	0.58&	1.2730&	1&	0.033\\ \hline
Yes&	4947&	2.24&	0.0915&	13&	0.362\\ \hline
No&	3185&	4.05&	0.3992&	4&	0.045\\ \hline
Yes&	4209&	6.63&	0.3742&	7&	0.084\\ \hline
Yes&	4444&	6.93&	0.3509&	21&	0.371\\ \hline

\end{tabular}%
\end{table}
%%%%%%%%%%%%%%%%%%%%%%%%%%%%%%%%%%%%%

Rupert et al. \cite{Rupter} used a logistic regression to predict the
probability of the debris-flow occurrence, and their results show that the
logistic regression is a valuable tool in the debris-flow prediction.
Therefore we utilize the logit model, again, to relate the binary outcome of
the debris-flows to five covariates listed in the last five columns in Table
3. Following the processing procedure demonstrated in Sec. 2.2, we out of 30
submodels found 3 submodels having the minimum entropy of \symbol{126}2.93
as shown in Table 4, i.e. Models 1, 2 and 3. The dilemma of entropy
resolution also appears in this case. Two approaches are useful in
determining the model(s) with the minimum entropy. We have listed in Table 4
all the calculated entropy of 30 submodels for the debris-flow data. The
calculated entropy of 30 submodels ranges between 2.9346 and 3.0726, and the
difference in entropy is about 0.138. When the resolution level in entropy
is assigned 10\% which is expected to be related to the measured precision
in observation Model 4 with the entropy of 2.9538 could then be
discriminated from the first 3 models with the entropy of \symbol{126}2.93,
because the entropy difference between Model 4 and the first 3 models is
larger than 10\% of 0.138. On the other hand, according to the debris-flow
data shown in Table 3, it seems fairly conservative to consider the
measurement precision is with three significant digits. Therefore, the
significant figure in entropy is also down to the second digits after the
decimal point and we still conclude the first 3 models with the minimum
entropy of \symbol{126}2.93 are the most preferable.

Same three variables of form factor (Column C in Table 4), numbers (Column D
in Table 4) and areas (Column E in Table 4) of landslides are incorporated
into all the 3 submodels with the entropy of \symbol{126}2.93, meaning that
those three variables are important to the debris-flow triggering,
particularly, in the studied areas of the dataset we used. We have noticed
that, in the two papers of \cite{LinCW00} and \cite{LinPS02}, the same
watershed was studied and both the observation spans of debris-flows are
after the 1996 Typhoon Herb. Therefore it is quite interesting to compare
our MEA result with the assessing variables used by the expert system in 
\cite{LinPS02}. In Lin et al. \cite{LinPS02} an overall debris-flow hazard
index is derived from a sophisticated GIS analysis of nine factors, i.e.
rock formation, fault length, landslide area, slope angle, slope aspect,
stream slope, watershed area, form factor and C factor (for the detailed
explanation of these factors, please refer to their paper). The data we
used, as mentioned above, only represents a relatively small dataset.
However, two variables of form factor and landslide area are agreeably
selected to be the important factors for the debris-flow triggering in both
our MEA result and the GIS analysis of Lin et al. \cite{LinPS02}, indicating
a fairly good performance of the MEA procedure. One important fact is that
our MEA procedure obviously provides a quantitative criterion in variable
selection for the debris-flow triggering while the reason for the selection
of those nine factors in Lin et al. \cite{LinPS02}, as they mentioned, is
quite subjective and based primarily on individual opinion and experience.

Then, the MEA procedure raises an open issue about whether the variable of
landslide number really does matter to the triggering of debris-flows. We
postpone to future work the examination of this issue.

%%%%%%%%%%%%%%%%%%%%%%%%%%%%%%%%%%%%%%%%%%%%%%%%%%%%%%%%%%%%%%%%
\vspace{-8pt} 
\begin{table}[h]
\caption{\baselineskip=8pt{\protect\footnotesize \textsf{Entropy (S) for total 
thirty submodels with different combinations of five variables in Table 3 
(A = Le, B = Ad, C = Ff, D = Nl, and E = Al). "1" or "0" denotes the variable 
selected or not selected in each submodel.}}}
\label{Table4}
\centering%
\begin{tabular}{|c|ccccc|c|}
\hline
Submodel No.&	A&	B&	C&	D&	E&	S\\ \hline
1&	0&	1&	1&      1&	1&	2.9346\\ \hline
2&	1&	0&	1&	1&	1&	2.9351\\ \hline
3&	0&	0&	1&	1&	1&	2.9355\\ \hline
4&	1&	1&	1&	0&	1&	2.9538\\ \hline
5&	1&	1&	0&	1&	1&	2.9542\\ \hline
6&	1&	0&	1&	0&	1&	2.9542\\ \hline
7&	1&	0&	0&	1&	1&	2.9557\\ \hline
8&	1&	0&	0&	0&	1&	2.9649\\ \hline
9&	1&	1&	0&	0&	1&	2.9653\\ \hline
10&	0&	1&	0&	1&	1&	2.9725\\ \hline
11&	0&	0&	0&	1&	1&	2.9738\\ \hline
12&	0&	1&	1&	0&	1&	2.9747\\ \hline
13&	0&	0&	1&	0&	1&	3.0065\\ \hline
14&	0&	1&	0&	0&	1&	3.0117\\ \hline
15&	1&	1&	1&	1&	0&      3.0240\\ \hline
16&     1&	0&	1&	1&	0&	3.0279\\ \hline
17&	0&	1&	1&	1&	0&	3.0289\\ \hline
18&	0&	0&	1&	1&	0&	3.0296\\ \hline
19&	1&	1&	1&	0&	0&	3.0446\\ \hline
20&	0&	1&	1&	0&	0&	3.0447\\ \hline
21&	0&	0&	0&	0&	1&	3.0447\\ \hline
22&	1&	0&	1&	0&	0&	3.0498\\ \hline
23&	1&	1&	0&	1&	0&	3.0550\\ \hline
24&	1&	0&	0&	1&	0&	3.0550\\ \hline
25&	0&	1&	0&	1&	0&	3.0560\\ \hline
26&	0&	0&	0&	1&	0&	3.0570\\ \hline
27&	0&	0&	1&	0&	0&	3.0607\\ \hline
28&	1&	1&	0&	0&	0&	3.0632\\ \hline
29&	1&	0&	0&	0&	0&	3.0638\\ \hline
30&	0&	1&	0&	0&	0&	3.0726\\ \hline
\end{tabular}
\end{table}
%%%%%%%%%%%%%%%%%%%%%%%%%%%%%%%%%%%

\section{Concluding remark}

In the data analysis, two questions are commonly addressed. What is the
pertinent model to best interpret experimental measurements for
understanding the physical system interested? And, what are the most
important variables that should be employed in the model? One may be able to
reveal natures and properties of the system through answering these two
questions. For the first question, unfortunately, there is no systematical
method to answer it. It is usually resolved through the methods of trials
and errors, empirical regressions, and some intuitive assumptions etc. We
therefore focus on answering the second question here. Our proposed MEA
procedure represents a systematical scheme to tackle this fundamental issue.
We establish, demonstrate and test our MEA procedure by two geoscientific
examples in this paper. The MEA procedure can then be satisfactory to
provide a quantitative criterion to the selection of relevant variables in
both examples.

To the course of data analysis, the MEA procedure seems simple and
straightforward. It is thus expected that the MEA procedure could be easily
extended to various geophysical/geological data analysis, where many
relevant or irrelevant possible measurements could obscure the understanding
of the highly complicated physical system. The triggering of debris-flows is
such an example. We would also like to emphasize here that the MEA procedure
only provides an honest way to extract the most effective information from
dazzling variables in hand. It can not guarantee the precision and the
correctness of measurement, which means the datum itself could be incorrect
and the measurement could be conducted in ill condition. This should be a
general property for all the data analyzing techniques.

\section*{Acknowledgment}
 CCC is grateful for research support from both
the National Science Council (ROC) and the Institute of Geophysics (NCU,
ROC). Research by CYT is grateful for research support from the 
National Science Council (ROC).

\end{document}